# Highlights

## Stochastic and Dynamic Fundamental Diagram for Mixed Traffic

Jiwan Jiang, Soyoung Ahn

- Develops a stochastic dynamic fundamental diagram for mixed traffic
- Integrates describing function analysis with Monte Carlo simulation
- Captures AV–HDV sequencing effects on traffic hysteresis

# Stochastic and Dynamic Fundamental Diagram for Mixed Traffic

Jiwan Jiang[a], Soyoung Ahn[b,*]

[a]*Zachry Department of Civil and Environmental Engineering, Texas A&M University, College Station, TX 77843, USA*
[b]*Department of Civil and Environmental Engineering, University of Wisconsin–Madison, Madison, WI 53706, USA*

**ABSTRACT**

This study develops a dynamic fundamental diagram (FD) framework tailored to mixed traffic environments comprising automated vehicles (AVs) and human-driven vehicles (HDVs). Describing function analysis is employed to derive approximate linear transfer functions for nonlinear HDV car-following models. A sequence-based stochastic dynamic FD is then formulated for mixed platoons, enabling the evaluation of hysteresis in the evolution of flow-density relations across different vehicle sequencing scenarios and AV penetration levels. Monte Carlo simulation results reveal that (i) differences in AV–HDV sequencing significantly alter the size of traffic hysteresis loops; and (ii) higher AV shares generally dampen hysteresis magnitude and variability, yet the net impact depends on how AVs are distributed within the platoon. The results suggest that traffic hysteresis in mixed environments is governed not only by the composition of AVs and HDVs, but also by how their interactions unfold through sequencing.

## 1. Introduction

The conventional approach to comprehending the properties of traffic flow has predominantly depended on aggregated metrics derived from a large volume of vehicles. These metrics encompass flow, density, and speed, with some being easily calibratable using data obtained from sparsely distributed sensors, offering fundamental insights into traffic performance. This approach gave rise to the formulation of the 'fundamental diagram' (FD), designed to depict traffic states in a 'steady-state', signifying the anticipated equilibrium flow concerning a specified speed or density (or conversely). The FD typically exhibits a distinct (linear or non-linear) correlation between flow and density in both uncongested and congested scenarios. Traditionally, the FD is constructed under an implicit steady-state assumption and is therefore static. It portrays a unique, deterministic flow–density (or speed–density) curve that is expected to hold once traffic has settled into equilibrium. However, detailed trajectory data reveal that real traffic rarely evolves along this curve when disturbed. Instead, the flow–density path typically traces an elliptical loop that encircles the equilibrium line—a phenomenon known as traffic hysteresis. Depending on whether the disturbance returns the system to the same equilibrium point or pushes it into a new congested state, hysteresis is commonly classified as oscillatory (Treiterer and Myers, 1974) or capacity-drop–induced (Zhang, 1999). These observations highlight a gap: static FDs are insufficient to explain the dynamic evolution of traffic states.

The advent of automated vehicles (AVs) is poised to reshape these dynamics, given that the behavioral patterns of AVs may differ fundamentally from those of traditional human-driven vehicles (HDVs). Empirical car-following campaigns show that commercial adaptive cruise control (ACC) systems adopt markedly different speed–spacing policies from human drivers and can both raise and lower capacity depending on controller settings (Makridis, Mattas, Anesiadou and Ciuffo, 2021). Micro- and meso-scopic simulations further suggest that even a modest AV penetration can dampen stop-and-go waves or, conversely, amplify them when poorly tuned (Shi, Zhou, Wu, Wang, Lin and Ran, 2021). Building on these observations, Jiang et al. analytically derived a dynamic fundamental diagram (FD) for homogeneous AV platoons by retaining higher-order CF terms, thereby explaining hysteresis in closed form (Jiang, Zhou, Wang and Ahn, 2024b; Jiang, 2025). Nevertheless, the literature still leaves several critical questions open.

First, most studies treat CF behavior as deterministic. This assumption ignores the intrinsic randomness of human reaction times and the growing empirical evidence that AV controllers themselves exhibit parametric uncertainty. Early stochastic work by Laval et al. reproduced HDV stop-and-go waves with a white-noise perturbation but did not relate those oscillations to the FD or to AV control variability (Laval, Toth and Zhou, 2014). More recent efforts inject stochasticity in different ways. Sun et al. extended three-phase theory with random slow-down probabilities to match rich congestion patterns, while Shiomi et al. used an LSTM model to output full acceleration distributions and identify variance spikes near merges and wave fronts (Sun, Zhang, Yuan, Tian and Wang, 2025; Shiomi, Li and Knoop, 2023). Nonetheless, all three studies remain HDV-centric and stop short of connecting microscopic randomness to a stochastic FD. Parallel work on approximate Bayesian computation and other data-driven calibration methods now quantifies parameter uncertainty for both HDVs and AVs (Jiang, Zhou, Wang and Ahn, 2024a; Jiang, Zhou, Jafarsalehi, Wang, Ahn and Lee, 2025); however, their implications for macroscopic flow–density relations are still unexplored.

---

[*]This research is supported by the U.S. National Science Foundation CMMI 2129765.
*Corresponding author
✉ sue.ahn@wisc.edu (S. Ahn)
ORCID(s): 0000-0003-4414-1959 (J. Jiang); 0000- (S. Ahn)

Second, mixed traffic that encompasses both AVs and HDVs adds another layer of complexity. Heterogeneous platoons combine (i) intra-class variability (different AV controllers from different manufacturers, diverse human driving styles) and (ii) inter-class interactions whose strength depends on the sequence in which AVs and HDVs appear (Qian, Li, Li, Zhang and Wang, 2017). Field observations indicate that the spatial distribution of AVs, clustered in platoons versus randomly dispersed, can substantially change macroscopic performance (Li, Chen, Zhou, Xie and Laval, 2022), yet most FD studies treat penetration rate, not sequence, as the sole descriptor of heterogeneity (Wang, Xu, Wu, Jiang and Yao, 2025; Zhou and Zhu, 2020). Multiclass continuum models with stochastic FDs have been proposed to acknowledge heterogeneity across vehicle classes (e.g.passenger cars, trucks)(Kim and Mahmassani, 2011). However, since their randomness is confined to class-averaged FD parameters rather than to driver-level interactions, they cannot capture sequence-dependent hysteresis or other transient effects.

Third, while macroscopic studies have assessed how aggregate AV penetration shifts capacity curves (e.g., CAV-dedicated-lane scenarios in (Chang and Chen, 2025)), the dynamic consequences—especially the way penetration modifies the size, orientation and persistence of hysteresis loops—remain largely unexplained.

In summary, we lack a comprehensive stochastic dynamic FD that (a) integrates high-order vehicle dynamics with stochastic behavior for both AVs and HDVs, (b) explicitly accounts for vehicle sequencing, and (c) quantifies how varying penetration rates reshape hysteresis. To bridge these gaps, we develop a sequence-based stochastic dynamic FD for mixed traffic (see **Figure 1**). The framework embeds two interacting layers of heterogeneity: i) Behavioral variability within and between vehicle classes (AV vs. HDV), modeled via probabilistic CF parameters; ii) Vehicle-order variability, captured through explicit enumeration of platoon sequences. Describing function analysis (DFA) linearizes each nonlinear CF model around sinusoidal disturbances, enabling closed-form expressions for first-order dynamics. Where closed forms become intractable, Monte-Carlo simulation completes the picture and recovers the full probability distribution of density and flow trajectories. The resulting framework reveals that (i) certain AV–HDV sequences suppress hysteresis more effectively than increases in uniform penetration, and (ii) increasing AV penetration reduces the magnitude and variability of hysteresis loops,confirming the stabilizing influence of well-tuned AV controllers in mixed traffic.

The remainder of the paper is structured as follows. Section 2 contrasts linear and nonlinear CF dynamics and motivates DFA. Section 3 details the DFA-based extraction of linear behavior from nonlinear CF laws. Section 4 formulates the proposed stochastic dynamic FD for mixed traffic. Section 5 presents numerical experiments and discusses the impact of sequencing and penetration on hysteresis. Section 6 concludes with managerial insights and future research directions.

## 2. Linear CF Behavior vs Nonlinear CF Behavior

A defining distinction between HDVs and AVs is the nature of their CF laws. Commercial AVs often rely on linear feedback controllers, commonly implemented through proportional–integral or spacing-policy formulations, whose response is characterized by a frequency-dependent transfer function (Jiang et al., 2025; Talebpour, Mahmassani, Hamdar et al., 2024). By contrast, the dominant HDV models, such as optimal velocity model (OVM) (Bando, Hasebe, Nakayama, Shibata and Sugiyama, 1995), generalized force model (GFM) (Helbing and Tilch, 1998), full velocity difference model (FVDM) (Jiang, Wu and Zhu, 2001), and intelligent driver model (IDM) (Kesting, Treiber and Helbing, 2010), are nonlinear.

**Figure 2** illustrates this contrast. Panels (a)–(d) show the two-dimensional acceleration surfaces $a = f(\Delta s_e, \Delta v)$ for the four HDV models, where the $x$-axis shows the relative speed and the $y$-axis indicates deviation from equilibrium spacing. The spacing between adjacent contour lines varies with the operating point, revealing an intrinsic nonlinearity. The non-uniform spacing between contour lines across the surface reflects the intrinsic nonlinearity of these models. In contrast, panels (e)–(f) illustrate lower-order (first-order) and higher-order AV controllers, whose uniformly spaced contours signify linear dynamic behavior.

For a linear CF law the small-signal gain from a sinusoidal disturbance in the predecessor's speed to the follower's speed is a scalar complex function

$$G(j\omega) = \frac{\Delta V_{\text{follower}}(j\omega)}{\Delta V_{\text{predecessor}}(j\omega)} = \frac{A\omega |G|}{A\omega} = |G| e^{j \angle G}., \tag{1}$$

which depends solely on the angular frequency $\omega$. Accordingly, one may employ classical Bode or Nyquist techniques.

In a nonlinear CF law the small-signal response depends on both $\omega$ and the disturbance amplitude $A$; the notion of a unique transfer function is unattainable. Exact frequency-domain analysis therefore becomes intractable.

## 3. Extraction of Linearized Dynamics via Describing Function Analysis

Describing-function analysis extends linear frequency-response ideas to mildly non-linear systems by retaining only the fundamental harmonic generated when a sinusoid of amplitude $B$ and frequency $\omega$ drives the non-linear block (Li and Zhou, 2025; Li, Wang and Ouyang, 2012; Zhou, Zhong, Chen, Ahn, Jiang and Jafarsalehi, 2023). Its validity rests on two standard assumptions: (i) High-frequency attenuation. The remainder of the feedback loop acts as a low-pass filter, therefore, higher harmonics are strongly damped. In CF applications this is plausible because driver perception–reaction delays ($\approx 1s$), power-train dynamics ($0.2-0.5s$), and ACC/AV controller bandwidths ($\leq 0.5Hz$) all act as first- or second-order lags. These lags strongly attenuate spectral content above $1\,rad\,s^{-1}$. (ii) Absence of sub-harmonic generation. The non-linear block does not create sub-harmonics. CF nonlinearities—saturation limits, dead zones, and $tanh$-type desired-speed functions—are static, single-valued, and memoryless; their Fourier series therefore contain only integer multiples of the excitation frequency. Sub-harmonics would require multivalued or time-varying elements, which are absent from the models analyzed here.

Under these conditions the output $y(t)$ may be approximated by its first Fourier harmonic:

$$y(t) = \sum_{k=0}^{\infty} Y_{k,1} \sin(\omega t) + Y_{k,2} \cos(\omega t) \approx Y_{1,1} \sin(\omega t) + Y_{1,2} \cos(\omega t), \tag{2}$$

where the coefficients of the first harmonics ($k = 1$), $Y_{1,1}$ and $Y_{1,2}$, are calculated by:

$$Y_{1,1} = \frac{1}{\pi} \int_0^{2\pi} y(t) \sin(\omega t)\, d(\omega t), \tag{3}$$

$$Y_{1,2} = \frac{1}{\pi} \int_0^{2\pi} y(t) \cos(\omega t)\, d(\omega t). \tag{4}$$

Building on the first harmonic approximation, the describing function serves as a complex-valued gain that characterizes the effect of the nonlinear element on a sinusoidal input. It captures both the amplitude scaling and phase shift introduced by the nonlinearity at the input frequency. The describing function is formally defined as:

$$N(B) = \frac{Y_{1,1} + jY_{1,2}}{B}, \tag{5}$$

The amplification and phase shift imposed on the sinusoidal input can be computed using $N(B) = \frac{\sqrt{Y_{1,1}+jY_{1,2}}}{B}$ and $\angle N(B) = \tan^{-1} \frac{Y_{1,2}}{Y_{1,1}}$.

In the following, we employ DFA to extract and approximate the linear dynamics of OVM, GFM, FVDM, and IDM, respectively. Due to the page limit, we takes OVM as an example, where we mathematically derive the describing functions and corresponding transfer functions for each CF law, providing a foundation for analyzing their frequency-domain characteristics. The derivation results for the rest CF models are shown in **Table 1** later.

### 3.1. Example: Optimal-Velocity Model (OVM)

The OVM prescribes the longitudinal acceleration of vehicle $l$ as

$$a_l(t) = \kappa \left[ V\left(\Delta x_l(t)\right) - v_l(t) \right], \tag{6}$$

where $\kappa = 2.0 \, (s^{-1})$ is the sensitivity parameter that implicitly captures the driver's response time. And the spacing $\Delta x_l(t) = p_l(t) - p_{l-1}(t)$. The nonlinear optimal velocity function $V\left(\Delta x_l(t)\right)$ is specified by the smooth, bounded hyperbolic-tangent form:

$$V\left(\Delta x_l(t)\right) = v_1 + v_2 \tanh\left[c_1\left(\Delta x_l(t) - c_2\right)\right], \tag{7}$$

where $v_1, v_2$ is the free-flow speed component, satisfying $v_1 + v_2 = v_{free}$. The parameter $c_1$ controls the steepness of the transition, and $c_2$ represents jam spacing. In equilibrium, the vehicles travel at constant speed $v_e$ with constant spacing $\Delta x_e$, satisfying $v_e = V(\Delta x_e)$. Thus, the equilibrium spacing can be derived as follows:

$$\Delta x_e = c_2 + \frac{1}{c_1} \text{arctan}\left(\frac{v_e - v_1}{v_2}\right). \tag{8}$$

Linearizing **Equation 6** around $(v_e, \Delta x_e)$ with $\Delta v_l(t) = v_l(t) - v_e$ and $\Delta x_l^*(t) = \Delta x_l(t) - \Delta x_e$ yields,

$$a_l(t) = \kappa \left[ V'(\Delta x_e) \cdot \Delta x_l^*(t) - \Delta v_l(t) \right]. \tag{9}$$

where $V'(\Delta x_e) = \frac{dV}{d\Delta x}\big|_{\Delta x_e}$, is the first-order approximation. Applying the Laplace transform (zero initial conditions) yields

$$G_{\text{OVM}}(s) = \frac{\Delta v_l(s)}{\Delta V(s)} = \frac{\kappa}{s + \kappa}. \tag{10}$$

For $s = j\omega$, the corresponding magnitude and phase of this transfer function are $\left|G_{OVM}(j\omega)\right| = \frac{\kappa}{\sqrt{\omega^2 + \kappa^2}}$ and $\measuredangle G_{OVM}(j\omega) = -\arctan\frac{\omega}{\kappa}$. Thus, in exponential form: $G_{OVM}(j\omega) = \frac{\kappa}{\sqrt{\omega^2+\kappa^2}} e^{-\arctan\frac{\omega}{\kappa}}$.

Then, assume the leader oscillates around equilibrium with amplitude $A$ and frequency $\omega$:

$$x_{l-1}(t) = v_e t + A \sin(\omega t), \tag{11}$$

The follower oscillates to respond with complex gain $G(j\omega) = |G| e^{j \measuredangle G}$:

$$x_l(t) = v_e t - \Delta x_e + A \left|G_{nl}(j\omega)\right| \sin\left(\omega t + \measuredangle G_{nl}(j\omega)\right), \tag{12}$$

where $\left|G_{nl}(j\omega)\right|$ and $\measuredangle G_{nl}(j\omega)$ are assumed nonlinear norm and phase shift. The reason to assume $G$ here is that the real analytical equation is extremely complex. Therefore, to be more intuitive, we have $G$ norm and $G$ angle assumed.

Letting the instantaneous spacing between $x_{l-1}(t)$ and $x_l(t)$ be represented as $\Delta x_l(t)$: $\Delta x_l(t) = A\sin(\omega t) - A\left|G_{nl}\right|\sin\left(\omega t + \measuredangle G_{nl}\right) = A_{\Delta x_l}\sin\left(\omega t + \theta_1\right)$. The resulting spacing oscillation amplitude is:

$$A_{\Delta x_l} = A\sqrt{1 + \left|G_{nl}\right|^2 - 2\left|G_{nl}\right|\cos \measuredangle G_{nl}}. \tag{13}$$

Consequently, the assumed oscillatory component of acceleration for OVM is:

$$a_l^{assd}(t) = -A\omega^2 \left|G_{nl}(j\omega)\right| \sin\left(\omega t + \angle G_{nl}(j\omega)\right). \tag{14}$$

Define $\Delta V(t) = V\left(\Delta x_e + \Delta x(t)\right) - V\left(\Delta x_e\right)$. For the sinusoidal spacing in **Equation 13**, the OVM describing function is

$$N_{\text{OVM}}\left(A_{\Delta x_l}, \omega\right) = \frac{1}{\pi A \Delta x_l} \int_0^{2\pi} \Delta V(\theta) \sin\theta \, d\theta. \tag{15}$$

In practice, this integral is computed numerically due to the nonlinearity of $V(\cdot)$. Using $G_{\text{OVM}}$ and $N_{\text{OVM}}$, the fundamental harmonic of the actual acceleration is

$$a_l^{DF}(t) = \kappa \left[A_{\Delta x} \cdot \left|N\left(A_{\Delta x}, \omega\right)\right| \sin\left(\omega t + \theta_2 + \angle N\left(A_{\Delta x}, \omega\right)\right) - A\omega \left|G\right| \cos\left(\omega t + \angle G\right)\right]. \tag{16}$$

For dynamic consistency, the fundamental harmonic of the acceleration obtained from describing-function analysis must equal the assumed acceleration, i.e., we want $a_l^{assd}(t) \approx a_l^{DF}(t)$. Equating the complex coefficients of the fundamental harmonic leads to Substituting Eq.13 yields the harmonic-balance equation

$$\left|G_{nl}\right| e^{j\angle G_{nl}} = G_{\text{OVM}}(j\omega) \, N_{\text{OVM}}(A_{\Delta x_l}, \omega) \frac{A_{\Delta x_l}}{A}. \tag{17}$$

Separating **Equation 17** into real and imaginary parts gives two nonlinear equations in $|G|$ and $\angle G$. Solving them numerically for each pair $(\omega, A)$ provides the amplitude ratio and phase lag that satisfy harmonic balance, completing the DFA for the OVM.

### 3.2. General Harmonic-Balance Procedure

In nonlinear systems, the input-output relationship is not solely determined by the frequency, as in linear systems, but also by the amplitude of the input signal. DFA provides a framework to estimate the gain and phase shift of the fundamental harmonic component when a sinusoidal input is applied. However, the describing function and transfer function alone are insufficient to fully determine the system response unless they satisfy a self-consistency condition. Otherwise the assumed sinusoidal response cannot persist. This is where the concept of harmonic balance becomes essential.

Thus, to fully characterize the frequency-domain behavior of nonlinear CF models, it is necessary to ensure that the assumed sinusoidal input and the resulting output oscillations are dynamically consistent. In the context of car-following models, the follower's acceleration response characterized by an amplitude gain and phase shift must align with the linear transfer function applied to the nonlinear describing function output. Mathematically, this harmonic balance condition is expressed as **Equation 17**. Algorithmically:

1. For a chosen pair $(A, \omega)$ compute $N_{\text{OVM}}(A_{\Delta x_l}, \omega)$ via **Equation 15**.
2. Substitute $G_{\text{OVM}}(j\omega)$ and $N_{\text{OVM}}$ into **Equation 17** to form one complex equation.
3. Numerically solve for the unknown magnitude $|G_{nl}(j\omega)|$ and phase $\angle G(j\omega)$.
4. Repeat over a grid of $(A, \omega)$ to map the amplitude-dependent frequency response.

## 4. Stochastic dynamic Fundamental Diagram

Section 3 provided amplitude–dependent transfer functions for CF laws under consideration. We now embed those results in a probabilistic framework that (i) propagates speed oscillations through a mixed AV/HDV platoon, (ii) maps the resulting spacing and speed fields to macroscopic density and flow, and (iii) yields a stochastic dynamic FD whose shape depends on controller heterogeneity, parameter randomness and vehicle sequence.

To derive the dynamic FD in a stochastic fashion, we first establish a probabilistic framework that characterizes the oscillation amplitude $A_l$ of the $l$-th vehicle in the platoon using a recursive amplitude model and sequence-based notation. The key idea is to propagate the initial oscillation amplitude introduced by the leading vehicle through the platoon, where each vehicle's response depends on its CF behavior and the amplitude received from its predecessor. This sequence-dependent propagation yields a stochastic process governing the evolution of oscillation amplitudes across the platoon. Building on this foundation, we derive the probability distribution functions (PDFs) of traffic density and flow over time by mapping vehicle-wise spacing and velocity oscillations into macroscopic traffic measures. The resulting dynamic FD captures not only the time-varying characteristics of traffic oscillations but also the impact of heterogeneity and randomness in CF behavior to provides a more comprehensive description of traffic dynamics in mixed platoons.

### 4.1. Probabilistic framework for a mixed platoon

Three sources of uncertainty arise once AVs and HDVs coexist. (1) Interaction accumulation: Even if all CF laws were linear, successive CF pairs generate distinct transfer functions that accumulate along the platoon. (2) Model heterogeneity: several (often nonlinear) CF laws coexist. Unlike linear models that depend only on frequency, nonlinear formulations exhibit amplitude-dependent gains and phases; therefore, their transfer functions vary with both input frequency and amplitude. (3) Parameter Randomness: CF model parameters are treated as random variables drawn from model-specific prior distributions, capturing inter-driver and inter-controller variability.

Consider a mixed platoon of length $N$ headed by a leading vehicle that oscillates at the principal frequency $\omega_p$ with amplitude $A_0$. Let $A_l$ be the speed-oscillation amplitude of the $l^{\text{th}}$ follower. Given the CF law index $Y_l \in \{1, \dots, I\}$ and its parameter vector $\Theta_l$, the DFA of Section 3 supplies an amplitude-dependent complex gain $G_l = G_l(Y_l, \Theta_l, A_{l-1}, \omega_p)$.

#### 4.1.1. Recursive Amplitude Model

To address the stochastic nature of heterogeneous CF behavior, we model the selection of CF laws and their associated parameter sets as random variables. Specifically, for the principal frequency $\omega_p$, each vehicle's oscillation amplitude is recursively computed using the following relationship derived from describing DFA:

$$A_l = |G_l(Y_l, \Theta_l, A_{l-1}, \omega_p)| \, A_{l-1}, \qquad l = 1, \dots, N, \tag{18}$$

with $A_0$ given. Since $A_l$ depends only on its predecessor and on the new randomness $(Y_l, \Theta_l)$, the sequence $\{A_l\}$ forms a time-homogeneous Markov chain whose one-step transition density is

$$f_{A_l \mid A_{l-1}}(a_l \mid a_{l-1}) = \sum_{i=1}^{I} p_i \int_{\Theta} \delta\!\left(a_l - |G_l(i, \vartheta, a_{l-1}, \omega_p)| \, a_{l-1}\right) D_i(\vartheta) \, d\vartheta, \tag{19}$$

where $p_i = Pr\{Y_l = i\}$ and $\delta(\cdot)$ denotes the Dirac delta function, enforcing the deterministic relationship between $a_l$ and $a_{l-1}$. $\vartheta$ denotes a generic parameter vector; and $D_i(\vartheta)$ is the distribution of $\Theta_l$ when $Y_l = i$. This recursive framework effectively captures the stochastic propagation of oscillations, incorporating the influences of both random CF law selection and parameter variability. Consequently, this probabilistic characterization of $A_l$ enables rigorous analysis of traffic stability in mixed platoon.

#### 4.1.2. Sequence based notation

Here introduce a sequence-based notation to systematically represent the mixed platoon configuration. Let $S = (i_1, i_2, \dots i_l)$, $l \in \{1, \dots, I\}$ denote the sequence list of CF laws within the platoon. For a fixed sequence $S$, define the joint parameter vector to capture the collection of parameters corresponding to each CF law: $\Theta_S = (\Theta_1, \dots, \Theta_I)^\top$, and $D(\Theta_S) = \prod_{j=1}^{N} D_{i_j}(\Theta_j)$, where $\Theta_s$ is the parameter set vector for sequence $S$, and $D(\Theta_s)$ is the CF law-specific parameter distribution vector for sequence $S$. Furthermore, denote the describing function gain and phase for an individual vehicle $l$ conditioned on sequence $S$ as: $g_l = |G_l(i_l, \theta_l, A_{l-1}, \omega_p)|$, and $\varphi_l = \angle G_l(i_l, \theta_l, A_{l-1}, \omega_p)$, where $g_l$ denotes the gain and $\varphi_l$ denotes the phase shift for vehicle $l$. It quantifies how each vehicle amplifies or dampens the oscillation. In DFA, each following vehicle, at the principal frequency $\omega_p$, exhibits a complex gain in the form $G_l(j\omega_p) = g_l e^{j\varphi_l}$. When a sinusoid input propagates through $l$ consecutive vehicles, the overall amplitude and phase shift become products and sums, respectively. Hence, we introduce order-sensitive cascade notations $P_l(S)$ and $\Theta_l(S)$ to represent the cumulative effects explicitly:

$$P_l(S) = \prod_{j=1}^{l} g_j, \qquad \Phi_l(S) = \sum_{j=1}^{l} \varphi_j, \qquad (20)$$

Therefore, the total amplitude and phase after $l$ followers are $A_l = P_l A_0$ and $\Phi_l$, respectively. Since $g_j$ and $\varphi_j$ depend on their position, any permutation of vehicles changes $P_l$ and $\Phi_l$; sequence therefore matters even when the penetration rate is fixed.

### 4.2. Stochastic Density and Flow

Mapping microscopic oscillations to macroscopic variables extends the deterministic derivation in (Jiang et al., 2024b) while retaining all random quantities. For a given sequence $S$ and associated parameter set $\Theta(S)$, the stochastic density and flow of an $N$-vehicle platoon are:

$$\tilde{k}(t, N \mid S) = \frac{N}{\sum_{l=1}^{N} \Delta x_{e,l} + \tilde{R}_N(S) A_0 \sin(\omega_p t + \phi_p - \tilde{\Phi}_N(S))}, \qquad (21)$$

$$\tilde{q}(t, N \mid S) = \frac{N v_e + A_0 \omega_p \sum_{l=1}^{N} P_l(S) \cos(\omega_p t + \phi_p + \Phi_l(S))}{\sum_{l=1}^{N} \Delta x_{e,l} + \tilde{R}_N(S) A_0 \sin(\omega_p t + \phi_p - \tilde{\Phi}_N(S))}, \qquad (22)$$

where the stochastic counterparts $\tilde{R}_N(S) = \sqrt{1 - 2P_N(S)\cos\Phi_N(S) + P_N^2(S)}$ and $\tilde{\Phi}_N(S) = \arctan\frac{P_N(S)\sin\Phi_N(S)}{1 - P_N(S)\cos\Phi_N(S)}$.

Since the gains $\{G_l\}$ are random, $\tilde{k}$ and $\tilde{q}$ admit no closed-form PDF; we therefore estimate their distributions via Monte-Carlo simulation by repeatedly sampling $(S, \Theta_S)$.

### 4.3. Probability distributions of density and flow

To determine the joint PDF of density and flow, we first calculate the conditional joint PDF for a given sequence $S$,

$$f_{\tilde{k},\tilde{q}\mid S}(k, q) = \int \delta\left(k - \tilde{k}\left(\vartheta_S\right)\right) \delta\left(q - \tilde{q}\left(\vartheta_S\right)\right) D\left(\vartheta_S\right) d\vartheta_S, \qquad (23)$$

where $D\left(\vartheta_S\right)$ is the joint PDF of CF parameters for sequence $S$ and $\vartheta_S$ denotes the parameter vector. By integrating over parameter distributions for each sequence, we obtain conditional PDFs. This integral enforces the mapping from parameters to the density-flow pair $(k, q)$.

Then marginalize these over all possible sequences by averaging the conditional PDFs weighted by their respective selection probabilities, resulting in the unconditional joint PDF:

$$f_{\tilde{k},\tilde{q}}(k, q) = \sum_{S \in \{1,\dots,6\}^N} \left(\prod_{l=1}^{N} p_{i_l}\right) f_{\tilde{k},\tilde{q}\mid S}(k, q), \qquad (24)$$

which defines the stochastic dynamic FD. $p_{i_l}$ indicates the probability of selecting the CF law for vehicle $i$.

**Equations 21–24** close the theoretical loop by linking microscopic randomness to macroscopic performance. Key insights emerging from the formulation are as follows. (1) Sequence sensitivity: For the same AV penetration rate, different spatial arrangements may widen or shrink the hysteresis loop—an effect quantified by the distribution of $\tilde{R}_N$ and $\tilde{\Phi}_N$. (2) Non-monotonic penetration effects: As $P_l$ and $\Phi_l$ are products and sums of random gains and phases, adding AVs does not guarantee smaller oscillations; poorly tuned controllers can increase $P_N$ in expectation. (3) Bridging micro and macro scales: The Markov-chain representation allows classical notions of string stability to be expressed as conditions on $P_l$, directly informing the variance of $\tilde{k}$ and $\tilde{q}$. These analytical expectations are validated and expanded through Monte-Carlo experiments in the following Section.

**Table 1**
DFA based Transfer Functions for HDVs

| CF Law | Acceleration Equation | Transfer Function |
|---|---|---|
| OVM | $a_l(t) = \kappa \left[ V\left(\Delta x_l(t)\right) - v_l(t) \right]$ | $\tilde{G}_{OVM}(s) = \frac{\kappa}{s+\kappa}$ |
| GFM | $a_l(t) = \kappa \left[ V\left(\Delta x_l(t)\right) - v_l(t) \right] + \lambda \, \Theta\left(-\Delta v_l(t)\right) \Delta v_l(t)$ | $\tilde{G}_{GFM}(s) = \frac{\kappa}{s+\kappa-\frac{\lambda}{2}}$ |
| FVDM | $a_l(t) = \frac{1}{\tau} \left[ V\left(\Delta x_l(t)\right) - v_l(t) \right] + \lambda \, \Delta v_l(t)$ | $\tilde{G}_{FVDM}(s) = \frac{\frac{1}{\tau}}{s+\left(\frac{1}{\tau}-\lambda\right)}$ |
| IDM | $a_l(t) = a_{max}\left[1 - \left(\frac{v_l(t)}{v_{max}}\right)^\delta - \left(\frac{s^*(v_l,\Delta v_l)}{s_l}\right)\right]^2$ | $\tilde{G}_{IDM}(s) = \frac{a_{max}}{j\omega}$ |

## 5. 5. EXPERIMENTAL RESULTS AND ANALYSIS

To evaluate the effectiveness of DFA in approximating linearized driving behaviors, we first investigate its performance in frequency response. Then we examine how CF laws, parameter uncertainty, and vehicle sequence shape the dynamic FD.

### 5.1. Frequency Response Analysis

Anchored in microscopic CF models and DFA, the framework connects individual driving behavior to macroscopic traffic patterns. We begin with a frequency-response study, evaluating the DFA-derived transfer functions of each CF law over the relevant spectrum. In the mixed traffic setting, AVs and HDVs follow distinct CF rules, and several alternative laws coexist within each class. The CF law index $Y_l \in \{1, ..., 6\}$ for vehicle $l$ is probabilistically drawn from the stochastic hybrid model of jiang2024generic. The hybrid model proceeds in three stages:

(1) Candidate pool. Four nonlinear HDV laws—FVDM, GFM, IDM, and OVM—constitute $Y_1$ to $Y_4$; two widely cited AV controllers—the lower-order linear (LL) feedback controller (Van Arem, Van Driel and Visser, 2006) and the higher-order linear (HL) controller (Zhou, Ahn, Wang and Hoogendoorn, 2020)—form $Y_5$ and $Y_6$.

(2) Massive sampling and ABC calibration. Millions of parameter particles are sampled from priors for every law and evaluated against NGSIM trajectories. A likelihood-free Approximate Bayesian Computation filter retains only those particles whose simulated errors fall below a universal threshold. The share of accepted particles associated with each law is interpreted as the posterior probability that the law generated the data.

(3) Posterior parameter sets. **Figure 3** displays the resulting HDV law probabilities, calibrated on NGSIM: $\beta_{\text{GFM}} = 0.352$, $\beta_{\text{FVDM}} = 0.311$, $\beta_{\text{IDM}} = 0.292$, $\beta_{\text{OVM}} = 0.044$.

Then, for every HDV follower we first sample

$$Pr(Y_l = i) = \beta_i \quad (i = 1, ..., 4) \tag{25}$$

and then draw its parameter vector $\Theta_l \sim D_{Y_l}(\Theta)$.

Besides, the transfer function $\tilde{G}$ for HDV follower is derived from the DFA as aforementioned. The table below summarizes the results.

For LL, the control system state at time $t$ is defined as $[\Delta x_e(t) - \Delta x_l(t), v_{l-1}(t) - v_l(t)]^\top$, where the first term is the deviation from equilibrium (desired) spacing and the second term is speed difference. Here, we incorporate two gains: spacing deviation feedback gain $k_s$ and speed deviation gain $k_v$. These gains are time-invariant and utilized to regulate the deviation from equilibrium spacing and the speed difference, respectively. Thus, the acceleration is given by:

$$a_l(t) = k_s \cdot \left(\Delta x_e(t) - \Delta x_l(t)\right) + k_v \cdot \left(v_{l-1}(t) - v_l(t)\right). \tag{26}$$

Then the corresponding transfer function is given as:

$$G_{LL,l}(j\omega) = \frac{k_s e^{-j\omega} + jk_v \omega e^{-j\omega}}{-\omega^2 + k_s e^{-j\omega} + j\left(k_v + k_s \tau\right) \omega e^{-j\omega}}. \tag{27}$$

The HL controller, $Y_6$, is defined in a state-space framework with state vector: $x_l(t) = [\Delta s_l(t), \Delta v_l(t), a_l(t)]^\top$, where three terms respectively represent deviation from the desired spacing, relative speed, and acceleration. The updated equation is given by $x_{l,t+1} = A_l x_{l,t} + B_l u_{l,t} + D_l a_{l-1,t}$, where $u_{l,t} = k_l x_{l,t} = [k_{s_l}, k_{v_l}, k_{a_l}] x_{l,t}$ is the control input (desired acceleration) and $a_{l-1,t}$ is the acceleration of the preceding vehicle (treated as an external input). $k_{s_l}$, $k_{v_l}$, and $k_{v_l}$ are gains on $\Delta s_l(t)$, $\Delta v_l(t)$ and $a_l(t)$, respectively.

Accordingly, the frequency-domain transfer function for HL is expressed as:

$$G_{HL,l}(j\omega) = \frac{k_{s_l} + jk_{v_l}\omega - k_{a_l}\omega^2}{-TT_l(j\omega)^3 + \left(k_{a_l} - 1\right)(j\omega)^2 + \left(TT_l k_{s_l} + k_{v_l}\right) j\omega + k_{s_l}}. \tag{28}$$

where $TT_l$ is the vehicle's actuation time constant.

Now that the CF law for each vehicle has been probabilistically determined, all prior distributions for each law are adjusted such that the equilibrium spacing aligns with a set equilibrium speed $v_e = 15 m/s$ (See Appendix Table 4). Notably, the equilibrium spacings for OVM and GFM are identical as shown in **Equation** 8. The equilibrium spacings for FVDM and IDM are respectively $\Delta x_{e-FVDM} = \text{lint}\left(\beta + \arctan\left(\frac{v_e - v_1}{v_2}\right)\right)$ and $\Delta x_{e-IDM} = \frac{s_0 + v_e T}{\sqrt{1-\left(\frac{v_e}{v_{max}}\right)^\delta}}$.

The equilibrium spacing for LL and HL uses the widely adopted constant time gap policy $\Delta x_e(t) = v_l(t) \times \tau + s_0$, where $\tau$ and $s_0$ represent constant desired time gap and standstill spacing, respectively. More details can be found in (Jiang, 2025).

**Figure** 4 condenses the frequency-response surfaces $|G(j\omega)| - G(j\omega)$ extracted from NGSIM I-80 data for $\omega \in [0.1, 0.4]\, rad/s$. **Figure** 4(a) illustrates the frequency response characteristics of the FVDM. Most selected particles have frequency response norms close to 1, indicative of marginal stability. **Figure** 4(b) illustrates the frequency responses for GFM. Similar to FVDM, most frequency response norms cluster around 1, indicating marginal stability. Under these circumstances. This behavior aligns closely with Newell's classical simplified car-following model (Newell, 2002), where vehicles exhibit neutrally stable following patterns with limited oscillation amplification or damping. The results of the OVM analysis in **Figure** 4(c) show that the spread of the norms widens, yet still close to 1, indicating that weak damping allows oscillations to persist. The results for IDM in **Figure** 4(d) exhibit substantial variability despite involving the fewest particles in the analysis. The considerable scatter and inconsistency imply IDM's high sensitivity to parameter choice and driver heterogeneity.

The analysis shows that stable behavior requires the describing function gain $|G|$ stay near but not depart markedly from 1. Across the tested frequency band, none of the four HDV models achieves robust string stability: as $\omega \to 0$, we consistently observe $|G| \to 1$, signaling marginal stability and a propensity for wave amplification. Of the models examined, IDM is the most structurally non-linear, displaying the widest spread in $|G|$ and, consequently, the greatest variability in dynamic response.

### 5.2. Stochastic Dynamic FD

We conduct a numerical experiment to examine the influence of AV penetration rate and the AV-HDV sequence on hysteresis on stochastic dynamic FD. For HDV followers, model parameters are calibrated using a newly collected high-resolution trajectory dataset collected at I-24 (Gloudemans, Wang, Ji, Zachar, Barbour, Hall, Cebelak, Smith and Work, 2023), recorded at 25 Hz, which provides detailed vehicle dynamics suitable for capturing microscopic traffic behavior. The data we used covers a 2.2 km freeway segment and spans the morning peak period from 6:30 to 7:00 AM, offering rich observations under congested and transitional traffic conditions. The high temporal resolution ensures accurate estimation of speed, acceleration, and spacing, making it particularly well-suited for stochastic and dynamic FD analysis. For AV followers, model parameters are trained using the CAR MODEL I subset of the MA dataset (Li et al., 2022), which contains controlled experiments of automated vehicle operations. This dataset includes AV platoon driving scenarios with different control settings and multiple consecutive AVs, enabling the capture of cooperative behaviors that differ from human driving. The uniform prior distribution ranges are included in **Appendix 7.1**.

The test platoon consists of 21 vehicles: one leader followed by 20 vehicles. The leading vehicle is driven by a sinusoidal oscillation of frequency $\omega = 0.1 rad/s$ and initial amplitude $A_0 = 10m$. Unless noted otherwise, the AV penetration rate is set to 50 %. Each simulation runs for 200 $s$, and Monte Carlo sampling is used to account for parameter uncertainty and vehicle-sequence permutations. From the results, we estimate the PDF of the terminal oscillation amplitude $A_N$, providing a robust statistical characterization of mixed-platoon dynamics.

### 5.2.1. Effects of Platoon Sequence

Define four representative vehicle sequence scenarios: Scenario 1 (AV First): all AVs are positioned before HDVs; Scenario 2 (HDV First): all HDVs precede the AVs; Scenario 3 (Alternating AV/HDV): AVs and HDVs alternate sequentially; and Scenario 4 (Random Order): AVs and HDVs are placed in random order. In every scenario, the specific CF law assigned to each vehicle follows the probability distributions learned from the hybrid model in **Figure 3**. Accordingly, the selection probabilities are $p_{1-4} = [0.2553, 0.4833, 0.2334, 0.0300]$ for HDVs and $p_{5-6} = [0.6617, 0.3383]$ for AVs.

Monte Carlo simulations were conducted to generate 1,000 samples for each platoon sequencing scenario and evaluate hysteresis evolution. To visualize the results, we retained the top 30 % of PDF values—i.e., those exceeding the 70th-percentile threshold—and enclosed the corresponding high-density points with a convex-hull algorithm. **Figure 5** shows the resulting light-green hulls (core oscillation region) and a blue curve denoting the most representative hysteresis loop. **Table 2** summarizes key hysteresis metrics—loop width, length, area, average wave speed, high-density hull area, flow and density standard deviations, average Euclidean distance to the center, and RMS distance—at the 10th, 30th, 50th, 70th, and 90th percentiles (definitions in (Jiang et al., 2024b)).

Results in **Figure 5** and **Table 2** demonstrate that vehicle sequencing has a substantial impact on the magnitude and variability of traffic oscillations. Among the four scenarios, the AV first scenario yields the largest high-density convex hull ($780.25 veh^2/km \cdot h$) and the highest variability of flow (Standard deviation of flow: $105.00 veh/hr$). These outcomes suggest that placing AVs at the front may intensify control-induced disturbances. This is because HDVs exhibit amplitude-dependent nonlinear behavior: their amplification gains increase with the magnitude of the input oscillation. When AVs lead the platoon, their deterministic controllers initiate structured oscillations that propagate upstream. As these oscillations grow in amplitude, trailing HDVs respond more aggressively and with increased behavioral variability, reflecting the inherent limitations of human response under dynamically unstable conditions.

By contrast, the HDV first scenario yields the smallest high-density convex hull ($619.34 veh^2/km \cdot h$) and the lowest overall variability. This is attributable to the fact that AV behaviors are relatively insensitive to oscillation amplitude, resulting in more consistent performance when positioned upstream. The Alternating and Random sequencing strategies produce intermediate outcomes, with convex hull areas of $675.18 veh^2/km \cdot h$ and $690.74 veh^2/km \cdot h$, respectively, and moderate levels of variability.

These differences stem from how instabilities are triggered and amplified in a mixed, string-unstable platoon. This combination of higher gain and increased variability at larger amplitudes produces wider, more erratic hysteresis loops and a broader stochastic FD. All loops are clockwise: during congestion build-up (density rising) the platoon traces the upper branch (higher flow), and during recovery (density falling) it follows the lower branch (lower flow for the same density) (Daganzo, 2011). This orientation is consistent with earlier findings in (Shi et al., 2021; Kontar, Li, Srivastava, Zhou, Chen and Ahn, 2021).

### 5.2.2. Effects of AV Penetration Rate

Different AV penetration rates within a mixed platoon also influence the characteristics of the dynamic FD. We examine penetration levels of 0%, 33.3%, 66.7%, and 100%, consistently applying the AV first sequencing. More results for HDV first and Random sequencing are included in **Appendix 7.2**.

**Figure 6** presents the probability density distributions of the mixed dynamic FD under varying AV penetration rates. **Table 3** reports key hysteresis metrics from the 10th to the 90th percentile.

With increasing AV penetration, three systematic trends emerge. First, the area of the high-density convex hull decreases substantially, from 861.82 to 694.34 $veh^2/(km \cdot h)$—indicating that traffic oscillations diminish in magnitude. Second, flow variability increases slightly first and then declines, as evidenced by a change in the standard deviation from 104.29 to 105.96 then to 72.19 $veh/hr$, reflecting more consistent and stable system behavior after centain amount of AVs were introduced in the system. Third, the dynamic FD becomes increasingly concentrated around its center, with the RMS distance contracting from 104.30 to 99.47 $veh/hr$, reflecting more stable operation. Taken together, these results imply that HDVs inject greater randomness, whereas AVs—even from different manufacturers—exhibit more structured and deterministic responses.

**Table 2**
Traffic hysteresis loop measures with respect to sequence.

| Scenario | Percentile | $k_{range}$ | $Q_{range}$ | Width | Length |
|---|---|---|---|---|---|
| **AV first** | | | | | |
| | 10 | 1.48 | 240.05 | 240.06 | 0.22 |
| | 30 | 1.90 | 258.77 | 258.78 | 0.67 |
| | 50 | 2.23 | 275.26 | 275.26 | 1.15 |
| | 70 | 2.52 | 296.74 | 296.75 | 1.59 |
| | 90 | 3.10 | 324.01 | 324.02 | 2.38 |
| High-density convex hull area: 780.25veh$^2$/(km·h) | | | | | |
| Standard deviation of density: 1.04veh/km | | | | | |
| Standard deviation of flow: 105.00veh/h | | | | | |
| Average Euclidean distance to center: 91.20 | | | | | |
| RMS Euclidean distance to center: 105.00 | | | | | |
| **HDV first** | | | | | |
| | 10 | 1.68 | 160.44 | 160.46 | 0.05 |
| | 30 | 2.13 | 179.94 | 179.96 | 0.19 |
| | 50 | 2.39 | 192.88 | 192.89 | 0.34 |
| | 70 | 2.68 | 207.78 | 207.79 | 0.56 |
| | 90 | 3.11 | 229.76 | 229.77 | 0.93 |
| High-density convex hull area: 619.34veh$^2$/(km·h) | | | | | |
| Standard deviation of density: 1.00veh/km | | | | | |
| Standard deviation of flow: 73.83veh/h | | | | | |
| Average Euclidean distance to center:63.08 | | | | | |
| RMS Euclidean distance to center:73.84 | | | | | |
| **Alternating AV/HDV** | | | | | |
| | 10 | 1.61 | 200.67 | 200.69 | 0.11 |
| | 30 | 1.99 | 221.64 | 221.65 | 0.39 |
| | 50 | 2.30 | 236.68 | 236.69 | 0.74 |
| | 70 | 2.61 | 251.60 | 251.61 | 1.17 |
| | 90 | 3.29 | 273.89 | 273.9 | 1.92 |
| High-density convex hull area: 875.18veh$^2$/(km·h) | | | | | |
| Standard deviation of density: 1.02veh/km | | | | | |
| Standard deviation of flow: 88.98veh/h | | | | | |
| Average Euclidean distance to center:76.93 | | | | | |
| RMS Euclidean distance to center: 88.99 | | | | | |
| | 10 | 1.62 | 194.66 | 194.68 | 0.11 |
| | 30 | 2.01 | 216.95 | 216.96 | 0.43 |
| | 50 | 2.32 | 234.20 | 234.21 | 0.74 |
| | 70 | 2.67 | 248.90 | 248.91 | 1.15 |
| | 90 | 3.13 | 271.73 | 271.74 | 1.78 |
| High-density convex hull area: 690.74veh$^2$/(km·h) | | | | | |
| Standard deviation of density: 1.03veh/km | | | | | |
| Standard deviation of flow: 86.92veh/h | | | | | |
| Average Euclidean distance to center:74.85 | | | | | |
| RMS Euclidean distance to center:86.93 | | | | | |

The hysteresis loop retains its clockwise orientation at all penetration levels, indicating a consistent phase relation between density and flow. These findings corroborate those of Zhong et al.(Zhong, Zhou, Ahn and Chen, 2024) and underscore that higher AV penetration can mitigate traffic instability and reduce the stochasticity in traffic dynamics.

## 6. Conclusion

This study analyzed a stochastic, sequence-based dynamic FD for mixed traffic, explicitly modeling two intertwined sources of heterogeneity: (1) the coexistence of HDVs and AVs, and (2) the presence of multiple CF laws within each vehicle type. The framework derived probabilistic density–flow relations analytically where possible, complemented by Monte Carlo sampling when closed-form solutions are infeasible. The results yield two key insights. First, differences in AV–HDV sequencing markedly alter the size of traffic hysteresis loops, underscoring the importance of platoon composition beyond mere penetration rate. Second, higher AV shares generally dampen hysteresis magnitude and variability, yet the net impact depends on how AVs are distributed within the platoon. Together, these findings highlight that traffic hysteresis in mixed traffic depends not just on the share of AVs and HDVs, but also on the specific sequence in which the two vehicle types interact. Effectively managing these dynamics, especially during transitional periods of AV adoption—requires careful consideration of both vehicle placement and behavioral heterogeneity. Accounting for these dimensions can inform lane-management, platoon-forming, and controller-tuning strategies during the transitional era of partial AV adoption.

**Table 3**
Traffic hysteresis measures with respect to AV penetration rate.

| Scenario | Percentile | $k_{\text{range}}$ | $Q_{\text{range}}$ | Width | Length |
|---|---|---|---|---|---|
| **0% AV** | | | | | |
| | 10 | 0.79 | 248.61 | 0.20 | 248.62 |
| | 30 | 0.89 | 262.87 | 0.34 | 262.87 |
| | 50 | 0.98 | 272.77 | 0.45 | 272.78 |
| | 70 | 1.08 | 282.93 | 0.59 | 282.93 |
| | 90 | 1.19 | 295.22 | 0.78 | 295.22 |
| High-density convex hull area: 861.82 veh$^2$/(km·h) | | | | | |
| Standard deviation f flow: 104.29 veh/h | | | | | |
| Standard deviation of density: 0.87 veh/km | | | | | |
| Average Euclidean distance to center: 89.61 | | | | | |
| RMS Euclidean distance to center: 104.3 | | | | | |
| **33.3% AV** | | | | | |
| | 10 | 1.24 | 249.21 | 0.13 | 249.22 |
| | 30 | 1.56 | 267.65 | 0.37 | 267.65 |
| | 50 | 1.77 | 282.44 | 0.63 | 284.44 |
| | 70 | 2.03 | 299.53 | 0.98 | 299.53 |
| | 90 | 2.45 | 330.79 | 1.61 | 330.79 |
| High-density convex hull area: 802.95 veh$^2$/(km·h) | | | | | |
| Standard deviation of flow: 105.96 veh/h | | | | | |
| Standard deviation of density: 0.93 veh/km | | | | | |
| Average Euclidean distance to center: 91.77 | | | | | |
| RMS Euclidean distance to center: 105.97 | | | | | |
| **66.7% AV** | | | | | |
| | 10 | 1.84 | 227.00 | 0.50 | 227.01 |
| | 30 | 2.38 | 245.45 | 1.16 | 245.46 |
| | 50 | 2.67 | 264.82 | 1.61 | 264.83 |
| | 70 | 3.05 | 282.70 | 2.20 | 282.71 |
| | 90 | 3.64 | 313.26 | 3.01 | 313.27 |
| High-density convex hull area: 801.75 veh$^2$/(km·h) | | | | | |
| Standard deviation of flow: 99.47 veh/h | | | | | |
| Standard deviation of density: 1.15 veh/km | | | | | |
| Average Euclidean distance to center: 86.43 | | | | | |
| RMS Euclidean distance to center: 99.47 | | | | | |
| **100% AV** | | | | | |
| | 10 | 2.88 | 147.87 | 0.97 | 147.91 |
| | 30 | 3.52 | 173.56 | 1.97 | 173.58 |
| | 50 | 3.94 | 195.92 | 2.70 | 195.94 |
| | 70 | 4.43 | 213.99 | 3.39 | 214.01 |
| | 90 | 5.04 | 237.73 | 4.33 | 237.74 |
| High-density convex hull area: 694.34 veh$^2$/(km·h) | | | | | |
| Standard deviation of flow: 72.19 veh/h | | | | | |
| Standard deviation of density: 1.46 veh/km | | | | | |
| Average Euclidean distance to center: 62.53 | | | | | |
| RMS Euclidean distance to center: 72.20 | | | | | |

The analysis employs DFA to approximate nonlinear CF dynamics, assuming that higher-order harmonics are negligible. While DFA offers a mathematically rigorous first-order approximation, future work should explore alternative or hybrid techniques—such as harmonic balance with higher modes or data-driven surrogate models—to capture mixed-traffic oscillations with greater fidelity.

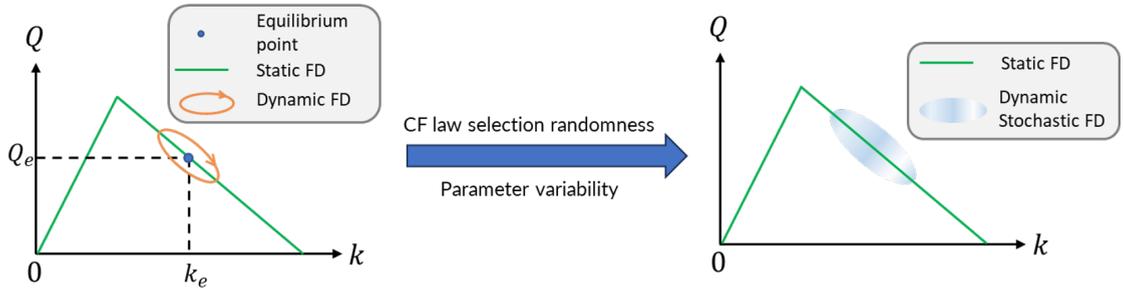

**Figure 1:** Dynamic FD vs. Stochastic Dynamic FD.

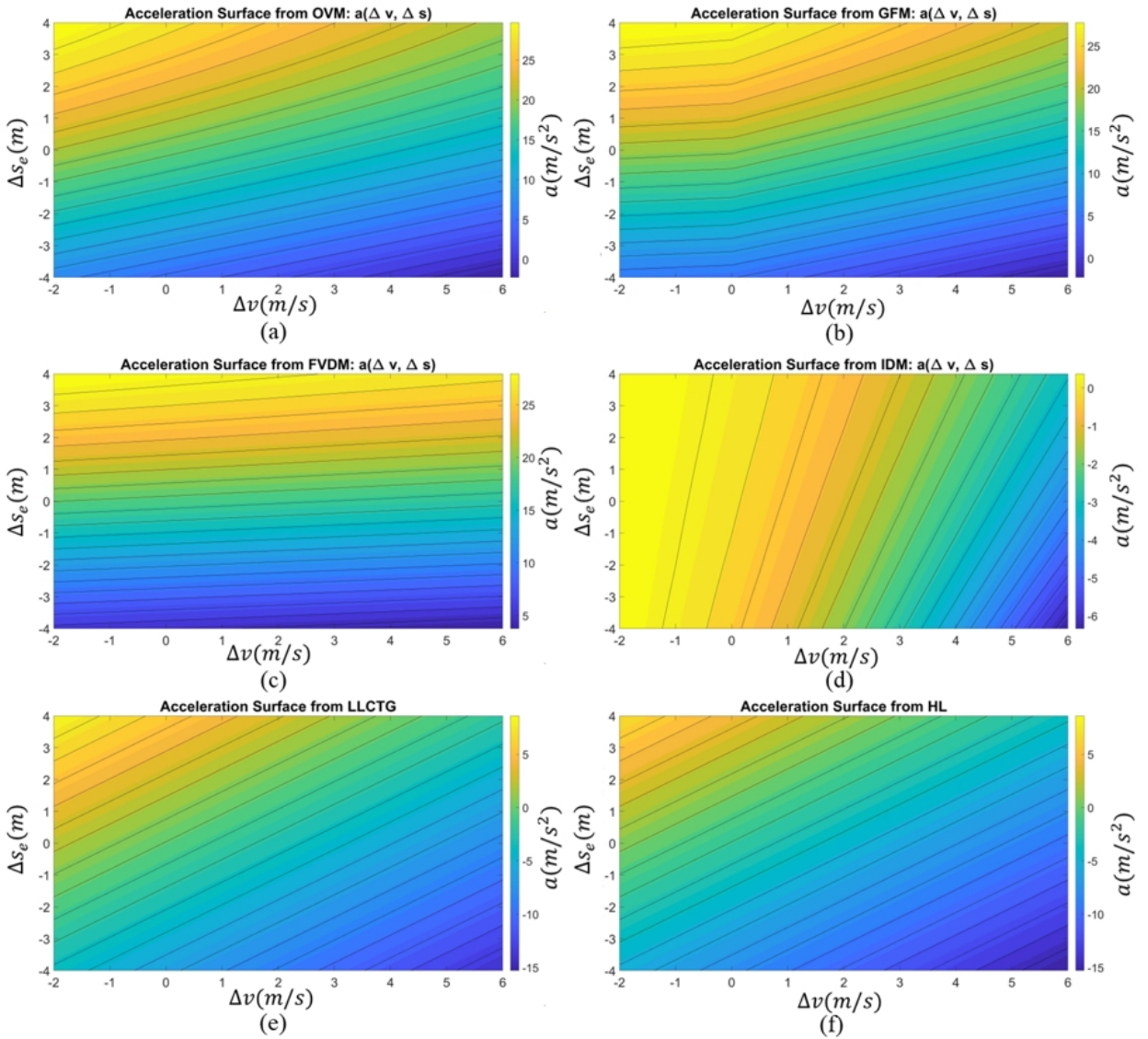

**Figure 2:** Comparison of Nonlinear Acceleration Behaviors of (a-d) HDV CF Laws; (e-f) Linear Feedback Controllers of CAV.

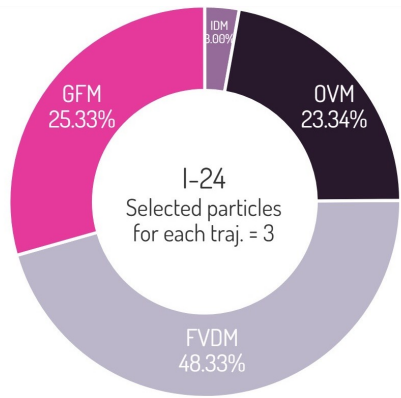

**Figure 3:** Hybrid Model Distribution Results ($N = 3$).

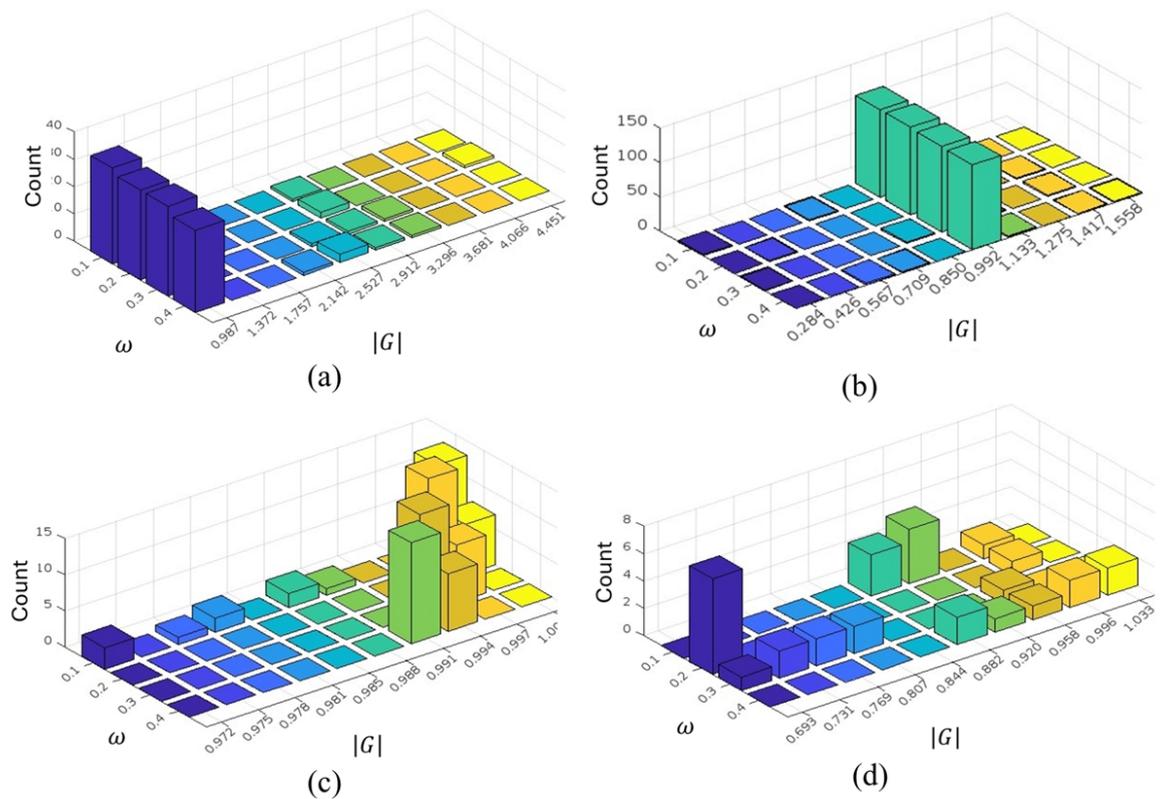

**Figure 4:** Frequency Response of G Using I-80 Dataset for (a) FVDM (b) GFM (c) OVM (d) IDM.

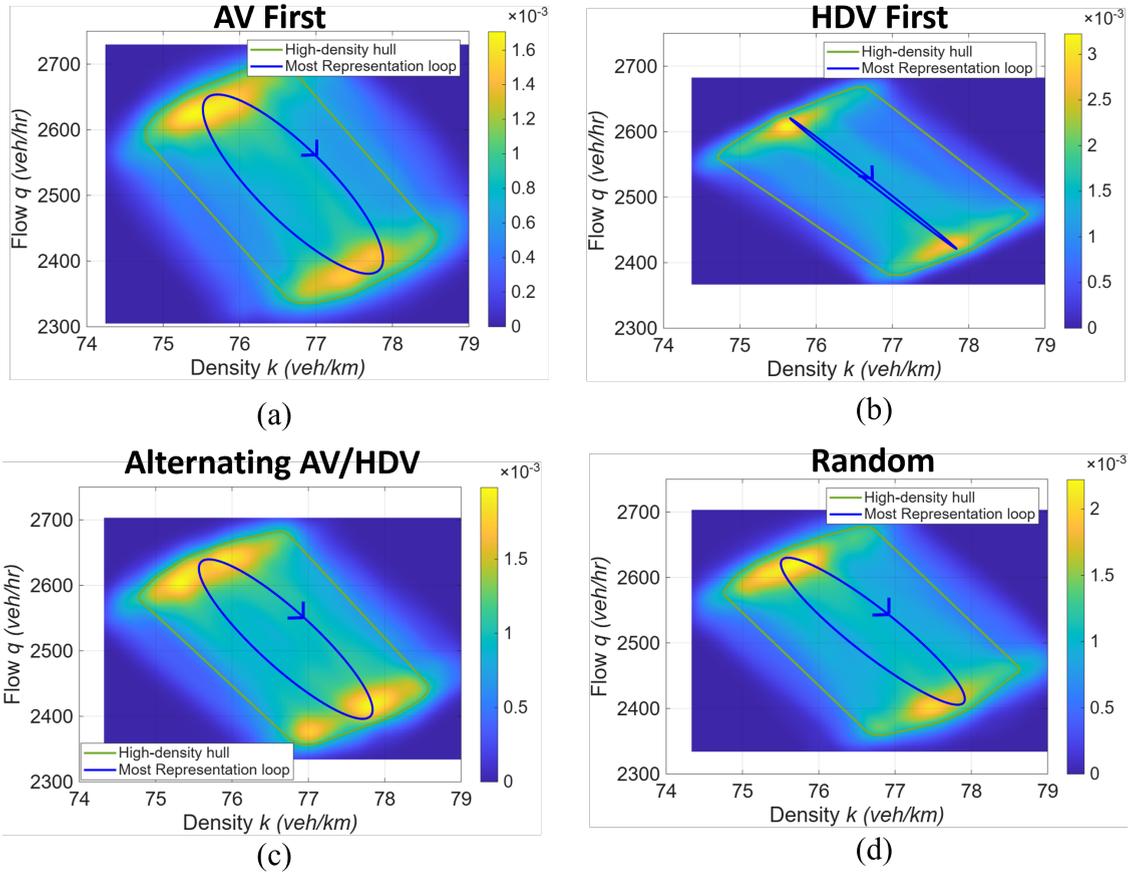

**Figure 5**: Probability Density Distribution of the Mixed Dynamic FD (Sequence) (a) AV first, (b) HDV first, (c) Alternating, (d) Random sequencing.

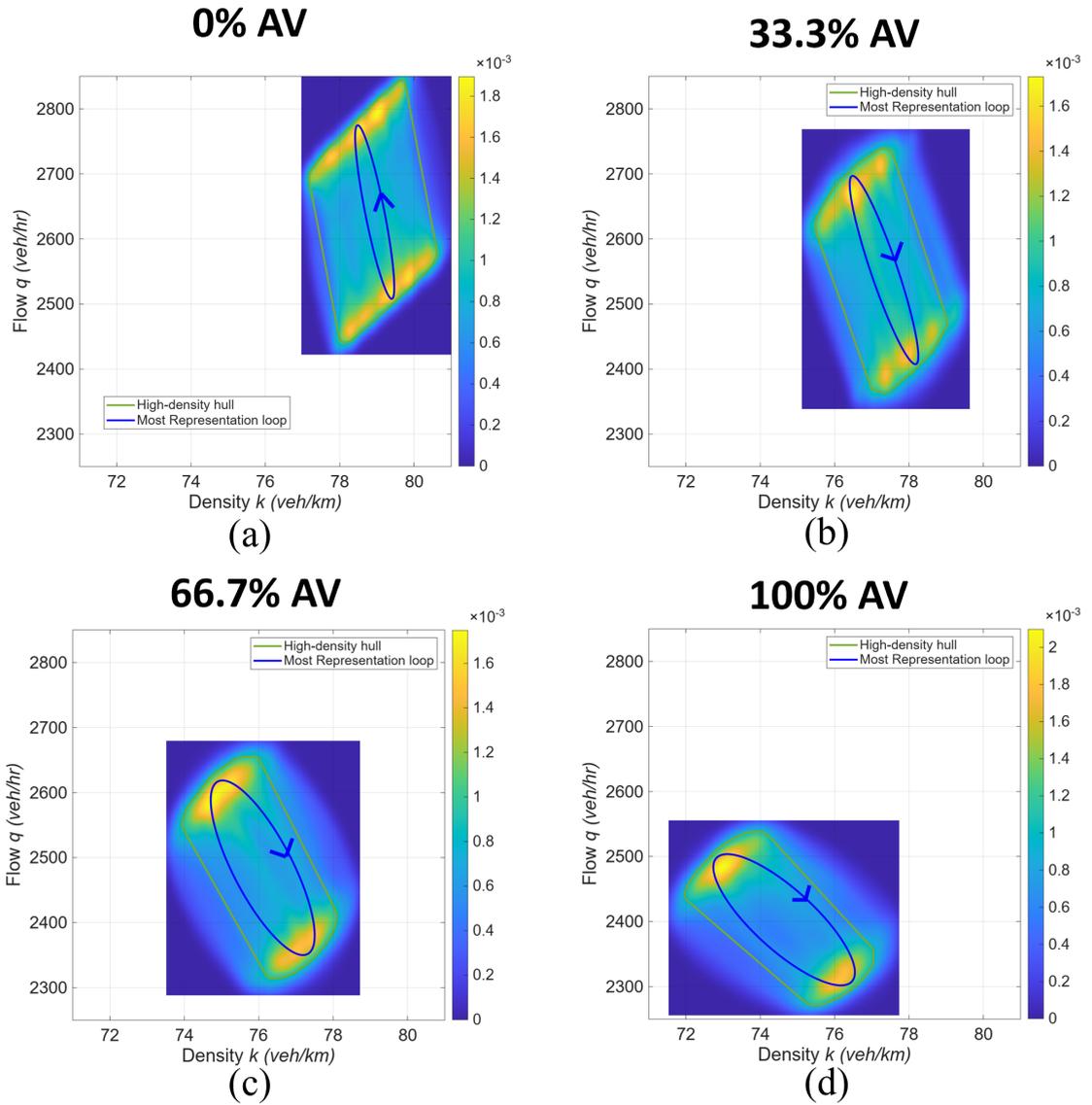

**Figure 6:** Probability Density Distribution of the Mixed Dynamic FD - AV first (Penetration Rate): (a) 0%AV, (b) 33.3%AV, (c) 66.7%AV, (d) 100%AV)

**Table 4**
Parameter Prior Ranges for CF Laws

| HL Parameters | Prior Ranges | LL Parameters | Prior Ranges |
|---|---|---|---|
| $\tau^*$ | [0.8, 1.2] | $\tau^*$ | [0.8, 1.2] |
| $TT$ | [0.1, 0.5] | $K_s$ | [0.1, 2.3] |
| $K_s$ | [0.1, 2.3] | $K_v$ | [0.1, 2.3] |
| $K_v$ | [0.1, 2.3] | $l$ | [3, 8] |
| $K_a$ | [−3, 0] | | |
| $l$ | [3, 8] | | |
| **GFM Parameters** | **Prior Ranges** | **OVM Parameters** | **Prior Ranges** |
| $\kappa$ | [1, 3] | $\kappa$ | [1, 3] |
| $\lambda$ | [2, 6] | $v_1$ | [10, 20] |
| $v_1$ | [10, 20] | $v_2$ | [15, 20] |
| $v_2$ | [15, 20] | $c_1$ | [0.1, 0.3] |
| $c_1$ | [0.1, 0.3] | $c_2$ | [15, 23] |
| $c_2$ | [15, 23] | | |
| **FVDM Parameters** | **Prior Ranges** | **IDM Parameters** | **Prior Ranges** |
| $\tau$ | [0.5, 2] | $a$ | [0.5, 2] |
| $\lambda$ | [1, 3] | $b$ | [1, 4] |
| $V_1$ | [10, 20] | $v_{\max}$ | [30, 50] |
| $V_2$ | [15, 20] | $T$ | [0.8, 1.5] |
| $l_{\text{int}}$ | [1.5, 1.7] | $s_0$ | [1, 3] |
| $\beta$ | [10, 15] | $\delta$ | [3, 4.5] |

## 7. Appendix
### 7.1. Appendix A: Parameter Prior Ranges for CF Laws
The prior ranges of parameters that we used for each CF law are shown in **Table 4**. We picked these based on empirical studies.

### 7.2. Appendix B: Different AV Penetration rates under HDV first sequence and Random sequence
To provide a more comprehensive analysis, we further consider AV penetration rates of 0%, 33.3%, 66.7%, and 100%, while consistently applying both HDV-first and random sequencing strategies. **Figure 7** and **Figure 8** present the corresponding results under HDV-first and random sequencing, respectively. While increasing AV penetration still reduces hysteresis magnitude and concentrates the dynamic FD, the rate of improvement is slower than in the AV-first scenario. In particular, the convex hull area remains larger and the probability density distribution is less compact at 33.3% and 66.7% AV, suggesting that sequencing plays a critical role beyond penetration rate alone.

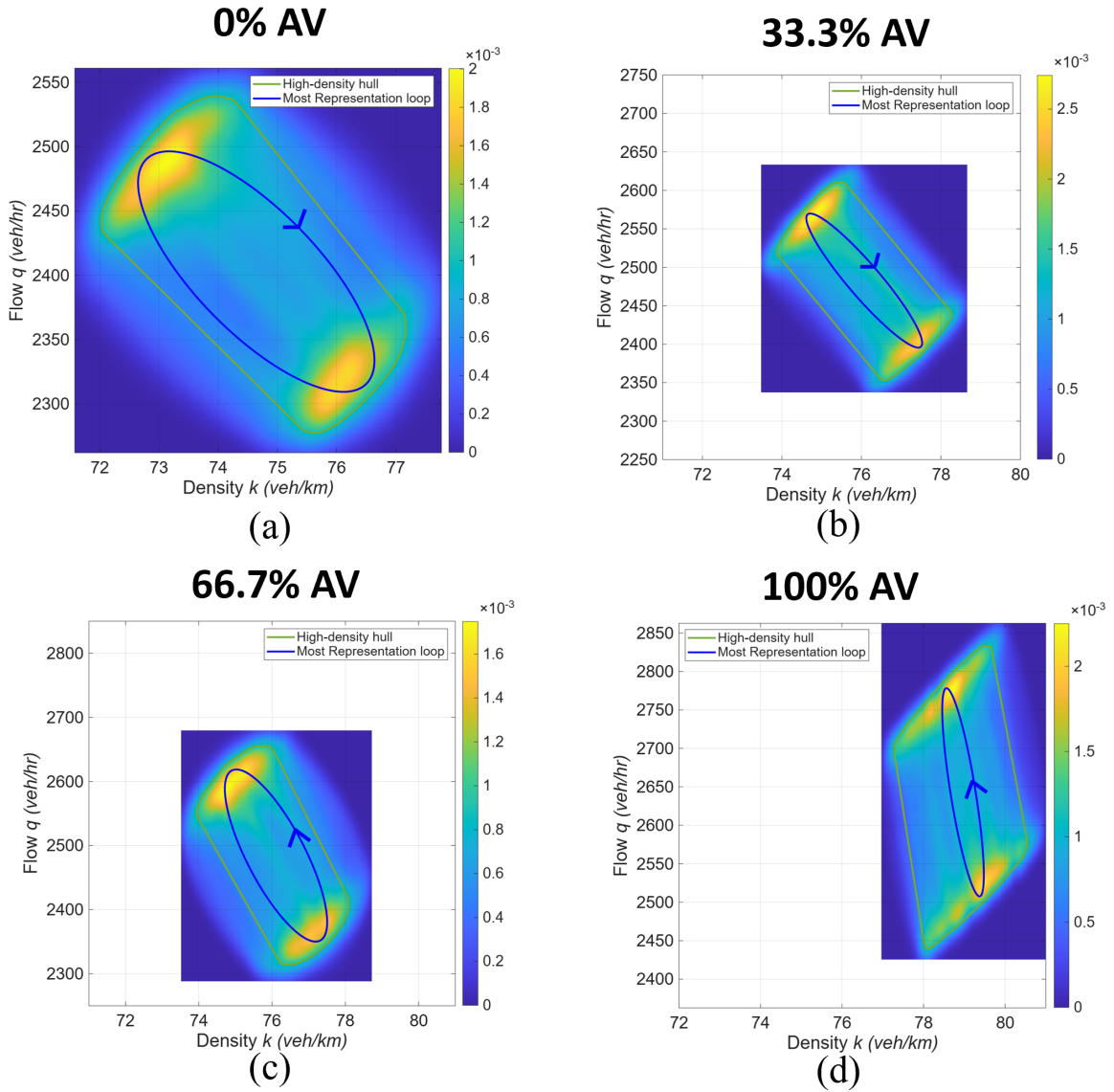

**Figure 7:** Probability Density Distribution of the Mixed Dynamic FD - HDV first (Penetration Rate): (a) 0% AV, (b) 33.3% AV, (c) 66.7% AV, (d) 100%AV)

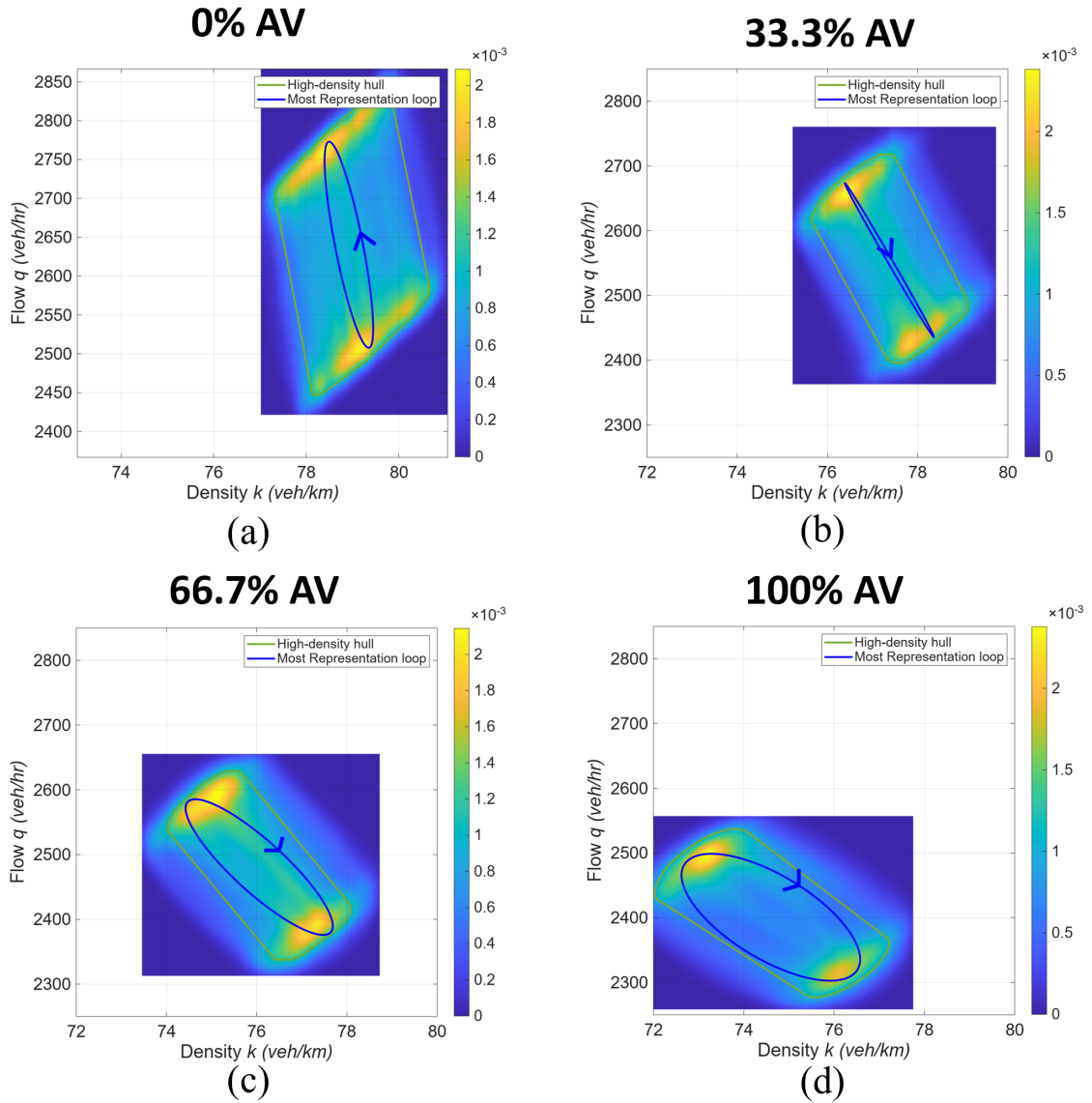

**Figure 8**: Probability Density Distribution of the Mixed Dynamic FD - Random (Penetration Rate): (a) 0% AV, (b) 33.3% AV, (c) 66.7% AV, (d) 100%AV)